\documentclass[12pt]{article}
\usepackage{graphicx}
\newcommand{\sss}{\scriptscriptstyle}

\newcommand {\be}{\begin{equation}} 
\newcommand{\ee}{\end{equation}}    
\def\ddt{\frac{\partial}{\partial t}}

\def\dds1{\frac{\partial}{\partial s_1}}

\def\vte{v_{{\sss T}e}}

\def\d{d\kern-0.8 ex\vrule height 1.3 ex depth-1.24 ex width 0.7 ex
\kern 0.15 ex}
\def\D{D\kern-1.7 ex\vrule height .87 ex depth-0.8 ex width 0.7 ex
\kern 0.95 ex}
\def\nabp{\nabla_{\perp}}

\begin{document}
\baselineskip 20 pt

\begin{center}

\Large{\bf Note on the role of friction-induced momentum conservation  in  the collisional drift wave instability}

\end{center}

\vspace{0.7cm}

\begin{center}

 J. Vranjes

{\em Center for Plasma Astrophysics, Celestijnenlaan 200B, 3001 Leuven,
 Belgium,
and
Facult\'{e} des Sciences Appliqu\'{e}es, avenue F.D. Roosevelt 50,
 1050 Bruxelles, Belgium}

\vspace{5mm}

S. Poedts

{\em Center for Plasma Astrophysics, Celestijnenlaan 200B, 3001 Leuven,
 Belgium, and Leuven Mathematical Modeling and Computational Science Center
 (LMCC)
}

\end{center}

\vspace{2cm}

{\bf Abstract:}
The collisional drift wave instability is reexamined taking into account the ion response in the direction parallel to the magnetic field lines, which appears  due to friction with electrons and which can not be omitted in view of the momentum conservation. A modified instability threshold is obtained. In plasmas with dominant electron collisions with neutrals, the instability threshold is shifted towards higher frequencies, compared to the case of dominant electron collisions with ions. The difference between the two cases vanishes when the ion sound response is negligible, i.e., when the instability  threshold disappears, and  both ions and neutrals react  to the electron friction in the same manner.

\vspace{2cm}


\vspace{2cm}

\pagebreak

\paragraph{I.} Within the standard fluid theory, the drift wave can be excited by  electron collisions.  In this case, the usual relation between the electron perturbed density and the perturbed potential, $n_1/n_0=e \phi_1/\kappa T_e$,  becomes modified due to the presence of the collisional term, so that the potential lags behind the density.$^{1,2}$ The effect appears regardless whether the electrons collide, in their predominant motion along the magnetic field vector, with ions or with neutrals.$^{3,4}$
Compared with the kinetic instability (which is due to the inverse electron Landau damping effect, that appears because  the  mode frequency is slightly below the diamagnetic frequency), the collisional instability is  dominant$^{5}$  provided  that the electron parallel mean-free path is smaller than the parallel wavelength. Hence  the interest in the drift modes with very large parallel wave-lengths and relatively short perpendicular wave-lengths.  In view of the early theoretical prediction$^{6,7}$ and experimental verification,$^{8}$ the amount of literature dealing with the drift wave is enormous.  More recently, collisional drift waves have been studied experimentally in cylindric plasma in Refs.~9-11. Details on the experimental investigation of global coherent structures associated with the drift mode in simple magnetized  torus can be found in recent Refs.~12-14.  The presence of charged grains on the drift wave in a cylindric configuration  has been studied in Ref.~15,  while the  role of the drift wave in the  transport phenomena may be found in the most recent Refs.~16, 17.

Some convenient approximations that are made in the derivations include the limit in which the complete  dynamics of the heavier particles (i.e.\ ions and neutrals) in the direction along the magnetic field is negligible, and the limit when electrons can be treated as inertia-less
\be
\vte \gg \omega/k_z\gg c_s,  \quad  c_s=(\kappa T_e/m_i)^{1/2}.
\label{e1}
\ee
In the presence of collisions, the friction force term in the electron parallel momentum equation is usually written in the form $-m_e n_e \nu_{ej} \vec v_e$, where $j=i, n$. As a result one obtains a standard  phase shift in the electron Boltzmann distribution $n_1/n_0=(e\phi_1/\kappa T_e)(1- i \delta)$, that is responsible for the mode growth.

However, the conservation of momentum implies that the friction term  in the electron momentum equation should read  $-m_e n_e \nu_{ej} (\vec v_e-\vec v_j)$ even when conventional criteria for a negligible parallel dynamics of heavier particles are fulfilled, so that the corresponding  momentum  component of the  heavier species includes the friction term $-m_j n_j \nu_{je} (\vec v_j-\vec v_e)$.  Below, we perform  derivations  with such a full friction force term  for some simple cases  in order to demonstrate the differences introduced by this friction-induced response of the heavier species.

\paragraph{II.} We first discuss {\em a plasma with dominant collisions between the charged particles}.  Note that this case may also include a rather weakly ionized plasma with $n_0\ll n_n$ (the index $0$ here and below  denotes the electron or ion equilibrium quantities).  This is   because of the  much larger cross section for collisions between charged particles. To have dominant collisions with protons in a  plasma containing electrons, protons and neutral atoms,  the electron number density  should satisfy the  condition
\be
\frac{n_0}{n_n}> \frac{3 \sigma_{en} (4 \pi \varepsilon_0 \kappa T_e)^2}{(8 \pi)^{1/2} e^4 L_{ei}}. \label{col}
\ee
Here we used the standard notation, and $L_{ei}$ is the Coulomb logarithm.  Taking as an example $T_e=10^4\;$K, which gives$^{18,19}$    $\sigma_{en}=2.5 \cdot 10^{-19}\;$m$^2$ (here $L_{ei}=6$ for $n_0=10^{18}\;$m$^{-3}$), it turns out that the electron-ion collisions are more frequent than the electron-neutral collisions  provided that $n_0/n_n> 0.009$. For the given temperature this is close to well known  estimate$^{20}$ showing that, in terms of electron collisions, an ion is equivalent to $3.4 \cdot 10^5 (300/T_e)^2\simeq 300$ neutral atoms.

We use  the  continuity equation for electrons and ions placed in an external magnetic field $\vec B_0=B_0\vec e_z$
\be
\frac{\partial n_j}{\partial t} + \nabla_\bot (n_j \vec v_{\bot j}) + \nabla_z (n_j  \vec v_{z j})=0, \quad j=e, i,  \label{ec}
\ee
where the linearized  perpendicular velocities of  electrons and ions are  given by
 \be
\vec v_{\bot e1}=\frac{1}{B_0}\vec e_z\times \nabla_\bot \phi_1 - \frac{\vte^2}{\Omega_e} \vec e_z \times \frac{\nabla_\bot n_{e1}}{n_{e0}}, \quad \vte^2= \kappa T_e/m_e, \label{epe}
\ee
\be
\vec v_{\bot i1}=\frac{1}{B_0}\vec e_z\times \nabla_\bot \phi_1 - \frac{1}{\Omega_iB_0}\ddt \nabp \phi_1. \label{epi}
\ee
Here, the effects of collisions on the perpendicular electron dynamics is neglected in contrast  to the parallel one, which is
justified$^{21}$  as long as $k_z^2 \Omega_e^2/(k_y^2 \nu_{ei}^2) >1$.

The electron parallel velocity is determined from
\be
0=e n_0\frac{\partial \phi_1}{\partial z} - \kappa T_e \frac{\partial n_1}{\partial z}- m_e n_0 \nu_{ei}(v_{ez1}-v_{iz1}), \label{e2}
\ee
and the ion velocity from
\be
\frac{\partial v_{iz1}}{\partial t}=- \frac{e}{m_i} \frac{\partial \phi_1}{\partial z} - \nu_{ie}(v_{iz1}-v_{ez1}). \label{e3b}
\ee
\paragraph{III.} In what  follows, we first present {\em the 'standard' derivation} in which the ion parallel velocity is only due to the parallel component of the perturbed electric field, while the friction induced term in (\ref{e3b})  is omitted with the usual excuse of  the huge difference in mass between the two species. In this case, the perturbed ion number density is described by$^{1,2}$
\be
\frac{n_1}{n_0} =\left(\frac{\omega_*}{\omega}  + \frac{k_z^2 c_s^2 }{\omega^2} - k_y^2 \rho_s^2 \right) \, \frac{e \phi_1}{\kappa T_e}. \label{e4}
\ee
Using Eqs.~(\ref{ec}), (\ref{e2}) the  perturbed electron number density can be written as
\be
\frac{n_1}{n_0}=\frac{\omega_* + k_z^2 c_s^2/\omega + i  k_z^2 D_z }{\omega + i  k_z^2 D_z } \, \frac{e \phi_1}{\kappa T_e},
\label{e5}
\ee
\[
D_z=\frac{\vte^2}{\nu_{ei}}, \quad \omega_*= -\frac{k_y \kappa T_e}{e B_0} \frac{n_0'}{n_0}.
\]
Here, we have taken $\nabla n_0=\vec e_x dn_0/dx$, the  perturbations are assumed to be of the form $\sim f(x) \exp(-i \omega t+ i k_y y + i k_z z)$, and we work in the frame  of a local approximation. The full collisional term, with perpendicular electron collisions included, should read
$i (k_z^2 \vte^2/\nu_{ei} + \rho_e^2 k_y^2 \nu_{ei})$, though the second term here is negligible in the limit discussed earlier. Here, $\rho_e=\vte/\Omega_e$ is the electron gyro-radius.

Note that $k_z^2 c_s^2/\omega$ in Eq.~(\ref{e5}), that  appears due to the term $ \nu_{ei}v_{iz1}$ in Eq.~(\ref{e2}), is usually omitted in standard derivations \cite{bel}.
The quasi-neutrality yields the dispersion equation
\be
\frac{\omega_*}{\omega}  + \frac{k_z^2 c_s^2 }{\omega^2} - k_y^2 \rho_s^2  =\frac{\omega_* + k_z^2 c_s^2/\omega + i  k_z^2 D_z }{\omega + i k_z^2 D_z }.
\label{e6}
\ee
We now take  the limit
\be
|\omega|\ll k_z^2 D_z, \label{e9}
\ee
and assume that $\omega$ and $\omega_*$ are of the same order. Used for convenience$^{1}$, the condition (\ref{e9})  is in fact  not always easily satisfied. Physically it describes$^{5}$ the condition of isothermal electrons along the field lines that has been assumed. It can be rewritten as $(\omega/k_z)/\vte\ll k_z \vte/\nu_{ei}$. The right-hand side gives the ratio of the electron mean free path $\vte/\nu_{ei}$ and the parallel wavelength, that in fact must be much less than unity in order to remain within a proper fluid theory (i.e., for collisions being able to maintain Maxwellian distribution).

A few comments here are noteworthy.
\begin{description}
\item{i)} If the parallel ion dynamics is {\em completely neglected}, we can write
  the dispersion equation as follows: $\Delta(\omega, k_y)\equiv \Delta_r(\omega, k_y) + i \Delta_i(\omega, k_y)\simeq \Delta_r(\omega_r, k_y) + i \omega_i \partial \Delta_r(\omega, k_y)/\partial \omega|_{\omega=\omega_r}   +i\Delta_i(\omega_r, k_y) =0$, where $|\Delta_i|\ll |\Delta_r|$ and where the subscripts $r, i$ denote 'real' and 'imaginary'. Setting  the real and imaginary  parts of this dispersion equation equal to zero  yields
  \be
  \omega_r\simeq \omega_* /(1+ k_y^2 \rho_s^2), \quad  \omega_i\simeq - \Delta_i(\omega_r, k_y)/(\partial \Delta_r/\partial \omega)_{\omega=\omega_r}
   = \frac{\omega_*^2}{k_z^2 D_z} \frac{k_y^2\rho_s^2}{(1+ k_y^2\rho_s^2)^3}. \label{e10}
  \ee
  The increment is proportional to $\nu_{ei}$, which implies that, with  the {\em fixed} ion background (in the parallel direction),   the more collisions the larger the growth of the wave is. Regarding the dependence on $k_y\rho_s$,  the increment is maximal at $k_y\rho_s=1/\sqrt{2}$.

\item{ii)} With the ion sound response taken into account (assuming that $k_z^2 c_s^2$ is same order or not much larger than $\omega^2$ and $\omega_* \omega$), the dispersion equation becomes
  \be
  \omega^2(1+ k_y^2 \rho_s^2) - \omega_* \omega - k_z^2 c_s^2 + \frac{i \omega}{k_z^2 D_z}\left[\omega(\omega- \omega_*) - k_z^2 c_s^2\right]=0. \label{e11}
  \ee
  Here, in the expansion on the right-hand side in Eq.~(\ref{e6}), the sound contribution yields a real and an imaginary term. The former, yielding a negligible correction to the real frequency, is neglected, and we have only kept the imaginary term. The real part of the frequency is determined from
  \be
  \Delta_r(\omega_r, k_y)\equiv \omega_r^2(1+ k_y^2 \rho_s^2) - \omega_* \omega - k_z^2 c_s^2=0. \label{e12}
  \ee
  This is used in the imaginary part, yielding
  \be
  \omega_i\simeq - \Delta_i(\omega_r, k_y)/(\partial \Delta_r/\partial \omega)_{\omega=\omega_r}=\frac{\omega_r^2}{k_z^2 D_z} \frac{\omega_r^2k_y^2\rho_s^2}{\omega_r^2(1+ k_y^2\rho_s^2) + k_z^2 c_s^2}. \label{e13}
  \ee
\item{iii)} We note that in fact in the standard literature \cite{bel} the sound contribution $k_z^2 c_s^2/\omega$ on the right-hand side in Eq. (\ref{e6}) is neglected. It is easily seen that the origin of this term is the term $m_e n_0\nu_{ei} v_{iz1}$ in Eq. (\ref{e2}). In this case instead of Eq. (\ref{e13}) we obtain
  \be
  \omega_i\simeq \frac{\omega_r^2}{k_z^2 D_z} \frac{\omega_r^2k_y^2\rho_s^2- k_z^2 c_s^2}{\omega_r^2(1+ k_y^2\rho_s^2) + k_z^2 c_s^2}. \label{e14}
  \ee
  Hence, there appears to be a threshold for the instability.
\end{description}

\paragraph{IV.}  Of course, a self-consistent analysis should include the full friction force effect in both, i.e., the  ion and electron, parallel equations. This is simply due to the fact that the two forces are necessarily equal by magnitude. Hence, we keep Eqs.~(\ref{e2}, \ref{e3b}) as they are,  {\em with the complete friction terms}.

Combining the two parallel momentum equations, and using the conservation of momentum $m_i\nu_{ie}=m_e \nu_{ei}$, yields $v_{iz1}= k_zc_s^2 n_1/(\omega n_0)$.
Instead of Eq.~(\ref{e4}), we now have
\be
\frac{n_1}{n_0}\left(1- \frac{k_z^2 c_s^2}{\omega^2}\right) =\left(\frac{\omega_*}{\omega}  - k_y^2 \rho_s^2 \right) \, \frac{e \phi_1}{\kappa T_e}. \label{e4a}
\ee
A procedure similar as earlier, now yields the electron number density
\be
\frac{n_1}{n_0}=\frac{\omega_* + i  k_z^2 D_z }{\omega - k_z^2 c_s^2/\omega + i  k_z^2 D_z } \, \frac{e \phi_1}{\kappa T_e},
\label{e5a}
\ee
instead of Eq.~(\ref{e5}). Compare these two equations with Eqs.~(\ref{e4}) and (\ref{e5}). Notice that in both Eqs.~(\ref{e4a}) and (\ref{e5a}), the ion parallel response appears in a completely different manner.

Within the same approximations as earlier (i.e., $\omega^2, \omega_* \omega, k_z^2 c_s^2 \ll k_z^4 D_z^2$), the dispersion equation that  we now have is:
\be
\omega^2(1+ k_y^2 \rho_s^2) - \omega_* \omega - k_z^2 c_s^2 + \frac{i (\omega^2- k_z^2 c_s^2)(\omega^2- \omega_* \omega
- k_z^2 c_s^2)}{\omega k_z^2 D_z}=0. \label{e15}
\ee
The real frequency is the same as in the earlier Eq.~(\ref{e12}). Hence, using Eq.~(\ref{e12}) the imaginary part of the frequency now becomes
\be
 \omega_i\simeq \frac{\omega_r^2 k_y^2 \rho_s^2}{k_z^2 D_z}\, \frac{\omega_r^2- k_z^2 c_s^2}{\omega_r^2(1+ k_y^2\rho_s^2) + k_z^2 c_s^2}=\frac{\omega_r^2 k_y^2 \rho_s^2}{k_z^2 D_z}\, \frac{\omega_* \omega_r- \omega_r^2 k_y^2 \rho_s^2}{\omega_r^2(1+ k_y^2\rho_s^2) + k_z^2 c_s^2}. \label{e16}
\ee
We remark the obvious difference in the instability threshold in the expression (\ref{e14}), and the correct expression (\ref{e16}), and we stress the absence of the threshold in Eq.~(\ref{e13}).

\paragraph{V.}
The ion response to the friction is usually neglected on the basis of the huge difference in mass of the different species. However, this difference in mass  may be compensated by the frequent electron collisions with ions, so that sooner or later the ions start to  move in the parallel  direction due to the electron drag.
To get a feeling on the effects of collisions and the corresponding time scales, we may discuss the following two separate cases.

\begin{description}
\item{a)} Assuming that the condition (\ref{e1}) is satisfied,  the ions respond in the parallel direction only through the friction. The electron velocity $V_0$  is  assumed  nearly constant by the parallel electric field of the wave.  From (\ref{e3b}), assuming that the ions are initially at rest, the ion velocity, normalized to $V_0$, becomes $1-\exp(-\nu_{ie} t)$.
   Taking  $n_0=10^{18}\;$m$^{-3}$, $T_e=10^4\;$K, $T_i= 2\cdot 10^3\;$K,  we have $\nu_{ie}=7 \cdot 10^3\;$Hz. A simple plot of the ion velocity reveals that it becomes close to 1 already after about 0.0007 seconds.

\item{b)} Taking another extreme case  where electrons initially, due to any external  reason acquire  a velocity $v_{e0}=V_0$, without any additional force,   and where $v_{i0}=0$. In this case the  electron velocity is not kept constant, the interaction of the two fluids yields the evolution of the two velocities:
    \be
    \vec v_e=\frac{\nu_{ie} \vec V_0}{\nu_{ei}+ \nu_{ie}} + \frac{ \nu_{ei} \vec V_0 }{\nu_{ei}+ \nu_{ie}}
    \cdot\exp[-(\nu_{ei} + \nu_{ie})t], \label{e17}
    \ee
    \be
    \vec v_i=\frac{\nu_{ie} \vec V_0}{\nu_{ei}+ \nu_{ie}} - \frac{ \nu_{ie} \vec V_0 }{\nu_{ei}+ \nu_{ie}}
    \cdot\exp[-(\nu_{ei} + \nu_{ie})t]. \label{e18}
    \ee
    Here, the electron and ion velocities monotonously change in time towards the common  velocity (the first term on the right-hand side)   $v_c\simeq V_0 m_e/m_i\ll V_0$ which, for the same parameters as above,  is achieved within the time interval shorter than $10^{-6}\;$sec. The characteristic time for the velocity relaxation is $\sim 1/(\nu_{ei} + \nu_{ie})$. This is presented in Fig.~1.
\end{description}

A real physical situation, as in the case of the drift wave discussed earlier,  is expected to be somewhere in between the two extremes presented above. Hence,  in spite of a huge mass difference, the collisions (friction) will force ions to move along the magnetic field lines, and,  due to the same reason,  the electron velocity amplitude associated with the drift wave is expected to be considerably smaller.

\paragraph{VI.} We shall check now {\em the case of a plasma with dominant electron collisions with neutrals}. The electron parallel momentum
Eq.~(\ref{e2}) now reads
\be
0=e n_0\frac{\partial \phi_1}{\partial z} - \kappa T_e \frac{\partial n_1}{\partial z}- m_e n_0 \nu_{en}(v_{ez1}-v_{nz1}). \label{e2a}
\ee
The ion dynamics is the same as above,  so we use Eq.~(\ref{e4}).
The dynamics of neutrals is completely described by
\be
\frac{\partial v_{nz1}}{\partial t}= - \nu_{ne}(v_{nz1}-v_{ez1}). \label{en}
\ee
This is used in Eq.~(\ref{e2a}), with the momentum conservation condition that  now reads $m_n n_n \nu_{ne}=m_e n_0 \nu_{en}$,
yielding
\be
v_{ez1}=\frac{i a k_z\vte^2}{\nu_{en}}\left(\frac{e \phi_1}{\kappa T_e}- \frac{n_1}{n_0}\right), \quad a=1+ \frac{i \epsilon \nu_{en}}{\omega}. \label{e19}
\ee
Here, $\epsilon=m_e n_0/(m_n n_n)$, and in view of Eq.~(\ref{col}) this is a small quantity for any plasma. For instance, for an electron-proton plasma in a hydrogen gas $m_n=m_i=m_p$, it is  of the order $10^{-6}$ or less. Equation~(\ref{e19}) is used in the electron continuity Eq.~(\ref{ec}) yielding
\be
\frac{n_1}{n_0}=\frac{\omega_* + i a k_z^2 D}{\omega  + i a k_z^2 D}, \quad D=\frac{\vte^2}{\nu_{en}}. \label{e20}
\ee
In the case  $\omega, \omega_* \ll k_z^2 D$,  Eq.~(\ref{e20}) can be written as
\be
\frac{n_1}{n_0}= \left[1+ \frac{\epsilon^2\nu_{en}^2}{\omega^2} - \frac{\epsilon \nu_{en}}{\omega} \frac{\omega + \omega_*}{k_z^2 D} +
\frac{i (\omega- \omega_*)}{k_z^2 D}\right]\left(1+ \frac{\epsilon^2\nu_{en}^2}{\omega^2} - \frac{2 \epsilon \nu_{en}}{k_z^2 D}\right)^{-1}.
\label{e21}
\ee
We have $1\gg 2 \epsilon \nu_{en}/k_z^2 D\gg \epsilon^2 \nu_{en}^2/\omega^2$, the real part  is therefore  very close to unity,  while  the imaginary part is simply $i (\omega-\omega_*)/k_z^2 D$. Consequently, combining  Eq.~(\ref{e21}) with Eq.~(\ref{e4}), the real part of the frequency appears described as earlier by Eq.~(\ref{e12}),  while the imaginary part becomes similar to Eq.~(\ref{e14}), viz.
\be
\omega_i\simeq \frac{\omega_r^2}{k_z^2 D} \frac{\omega_r^2k_y^2\rho_s^2- k_z^2 c_s^2}{\omega_r^2(1+ k_y^2\rho_s^2) + k_z^2 c_s^2}. \label{e22}
\ee
However, we stress again that Eq.~(\ref{e14}) follows from a formally  incorrect procedure.

Compared to the previously discussed $e-i$ collision case (\ref{e16}), it is seen that i)~in both cases  the threshold is caused by  the ion sound response, however,  ii)~the instability threshold in Eq.~(\ref{e22}) is shifted  towards higher frequencies (because $k_y \rho_s$ is usually less than  unity).

Although the two cases describe two physically different plasma environments, the explanation for case ii) should be as follows. The lower threshold frequency in the $e-i$ case implies that electrons experience a larger amount of collisions with ions within a wave period,  which is in fact  necessary to compensate for the ion movement  in the parallel direction (due to the parallel electric field). This is because moving ions (in the same direction as electrons) represent a less efficient  barrier for electron parallel motion and, in order to have the necessary  phase shift between the density and potential, the electrons should have more collisions for the instability to take place. On the other hand, in the $e-n$ case, a higher frequency (equivalent to a smaller amount of collisions) for the instability to develop  is possible because neutrals are less movable in the parallel direction (they do not react to the parallel electric field), and therefore they represent a more effective barrier.

\paragraph{VII.} To conclude,  the self-consistent  inclusion of the momentum conservation  in the ion and electron equations, which originates from the collisions between the two fluids, yields a different instability threshold that, to the best of our knowledge,  has not been discussed in the literature so far. The correct expressions (\ref{e16}) and (\ref{e22})  should be used for estimates of the growth rate and the instability threshold for the collisional drift mode.

\vspace{2cm}

\paragraph{Acknowledgements:}
The  results presented here  are  obtained in the framework of the
projects G.0304.07 (FWO-Vlaanderen), C~90205 (Prodex~9),  GOA/2004/01
(K.U.Leuven),  and the Interuniversity Attraction Poles Programme -
 Belgian State - Belgian Science Policy. JV would like to thank N. D'Angelo for fruitful discussions.

\vfill

\pagebreak

\vfill

\pagebreak

\noindent {\bf Figure caption:}

\vspace{1cm}

\noindent  {\bf Fig. 1.}  The evolution of the ion  velocity from Eqs.~(\ref{e17}, \ref{e18}) for an electron-ion plasma, normalized  to $\nu_{ie}  V_0/(\nu_{ei}+ \nu_{ie})$, and for $v_{e0}=V_0$, $v_{i0}=0$.

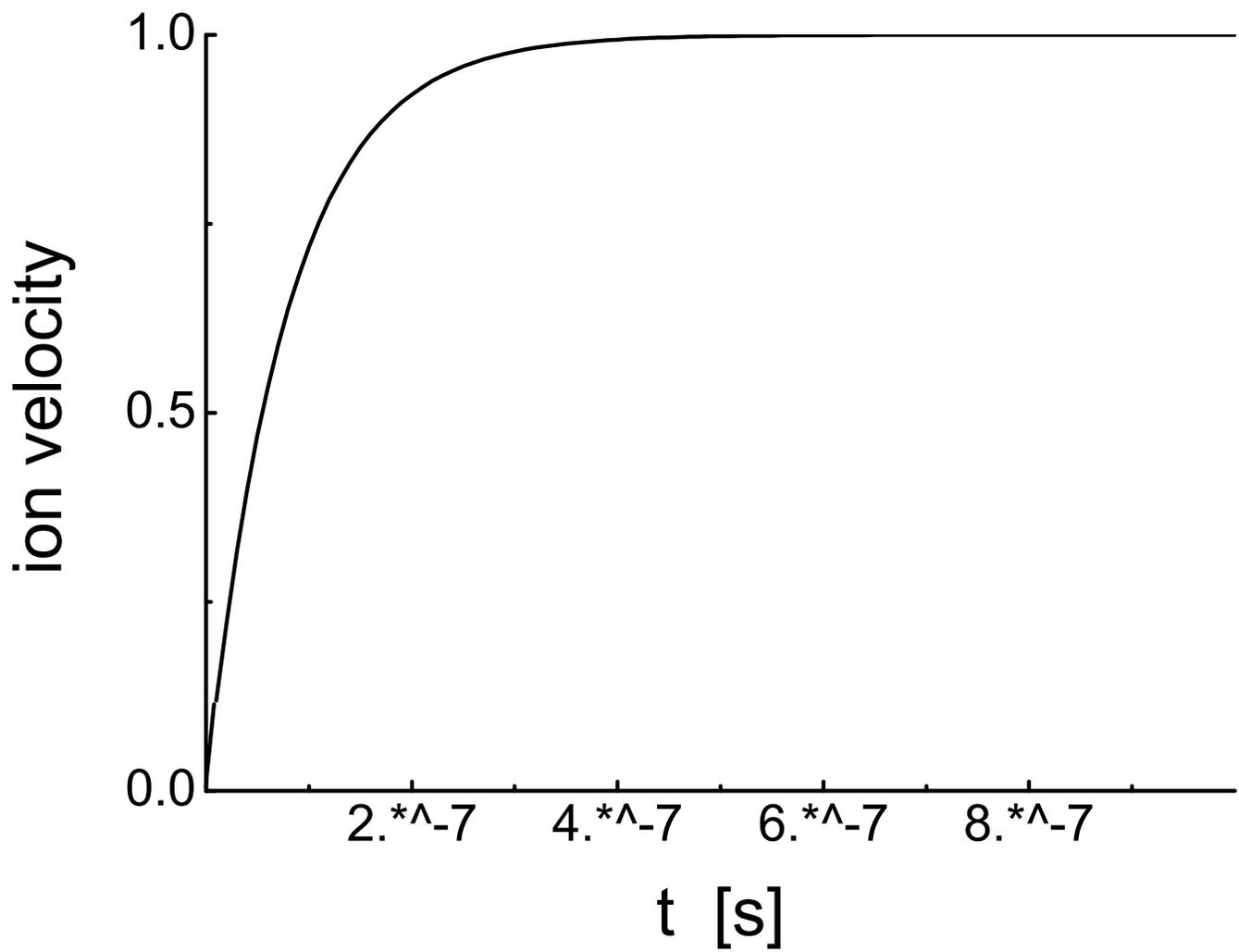
\end{document}